\documentclass[sigconf,screen,natbib]{acmart}

\usepackage{tabularx}
\usepackage{booktabs}
\usepackage{subcaption}
\usepackage[skip=0pt]{caption}

\usepackage{pgf, tikz}
\usetikzlibrary{arrows,automata,shapes}
\usetikzlibrary{positioning}
 \usepackage[noend]{algpseudocode}

\usepackage[inline]{enumitem}
\usepackage{amsmath}
\usepackage{graphicx}
\usepackage{makecell}

\makeatletter
\newif\if@restonecol
\makeatother

\usepackage[ruled,vlined]{algorithm2e}

\usepackage{color}

\newcommand{\tabincell}[2]{\begin{tabular}{@{}#1@{}}#2\end{tabular}}

\copyrightyear{2020}
\acmYear{2020}
\setcopyright{acmlicensed}
\acmConference[SIGIR '20]{Proceedings of the 43rd International ACM SIGIR Conference on Research and Development in Information Retrieval}{July 25--30, 2020}{Virtual Event, China}
\acmBooktitle{Proceedings of the 43rd International ACM SIGIR Conference on Research and Development in Information Retrieval (SIGIR '20), July 25--30, 2020, Virtual Event, China}
\acmPrice{15.00}
\acmDOI{10.1145/3397271.3401130}
\acmISBN{978-1-4503-8016-4/20/07}

\newcommand{\AuthorSpace}{ \ \ \ \ \ \ }

\author{
Nikos Voskarides$^1$\AuthorSpace
Dan Li$^1$\AuthorSpace
Pengjie Ren$^1$\AuthorSpace
Evangelos Kanoulas$^1$\AuthorSpace
Maarten de Rijke$^{1,\, 2}$
}
\affiliation{%
$^1$University of Amsterdam, Amsterdam, The Netherlands\quad
$^2$Ahold Delhaize, Zaandam, The Netherlands\\
nickvosk@gmail.com, d.li@uva.nl, p.ren@uva.nl, e.kanoulas@uva.nl, m.derijke@uva.nl
}

\renewcommand{\shortauthors}{Voskarides et al.}

\acmSubmissionID{233}

\begin{document}

\title{Query~Resolution~for~Conversational~Search with~Limited~Supervision}

\begin{abstract}

In this work we focus on multi-turn passage retrieval as a crucial component of conversational search.
One of the key challenges in multi-turn passage retrieval comes from the fact that the current turn query is often underspecified due to zero anaphora, topic change, or topic return.
Context from the conversational history can be used to arrive at a better expression of the current turn query, defined as the task of query resolution.
In this paper, we model the query resolution task as a binary term classification problem: for each term appearing in the previous turns of the conversation decide whether to add it to the current turn query or not.
We propose QuReTeC (\textbf{Qu}ery \textbf{Re}solution by \textbf{Te}rm \textbf{C}lassification), a neural query resolution model based on bidirectional transformers.
We propose a distant supervision method to automatically generate training data by using query-passage relevance labels.
Such labels are often readily available in a collection either as human annotations or inferred from user interactions.
We show that QuReTeC outperforms state-of-the-art models, and furthermore, that our distant supervision method can be used to substantially reduce the amount of human-curated data required to train QuReTeC.
We incorporate QuReTeC in a multi-turn, multi-stage passage retrieval architecture and demonstrate its effectiveness on the TREC CAsT dataset.
\end{abstract}


\let\backupauthors\authors
\renewcommand{\authors}{Nikos Voskarides, Dan Li, Pengjie Ren, Evangelos Kanoulas, and Maarten de Rijke}

\maketitle

\section{Introduction}
%
%
Conversational AI deals with developing dialogue systems that enable interactive knowledge gathering~\cite{gao_neural_2018}.
A large portion of work in this area has focused on building dialogue systems that are capable of engaging with the user through chit-chat~\cite{li_diversity-promoting_2016} or helping the user complete small well-specified tasks~\cite{peng_deep_2018}.
In order to improve the capability of such systems to engage in complex information seeking conversations~\cite{qu_attentive_2019}, researchers have proposed information seeking tasks such as conversational question answering (QA) over simple contexts, such as a single-paragraph text~\cite{choi_quac:_2018,reddy_coqa_2019}. 
%
In contrast to conversational QA over simple contexts, in conversational search, a user aims to interactively find information stored in a large document collection~\cite{culpepper_research_2018}. 

In this paper, we study multi-turn passage retrieval as an instance of conversational search: given the conversation history (the previous turns) and the current turn query, we aim to retrieve passage-length texts that satisfy the user's underlying information need~\cite{cast2019}.
Here, the current turn query may be under-specified and thus, we need to take into account context from the conversation history to arrive at a better expression of the current turn query.
Thus, we need to perform \emph{query resolution}, that is, add missing context from the conversation history to the current turn query, if needed.
An example of an under-specified query can be seen in Table~\ref{tab:quac_paragraph1}, turn \#4, for which the gold standard query resolution is: ``\emph{when was saosin 's first album released?}''.
In this example, context from all turns \#1 (``saosin''), \#2 (``band'') and \#3 (``first'') have to be taken into account to arrive to the query resolution.

Designing automatic query resolution systems is challenging because of phenomena such as zero anaphora, topic change and topic return, which are prominent in information seeking conversations~\cite{yatskar_qualitative_2019}.
These phenomena are not easy to capture with standard NLP tools (e.g., coreference resolution). 
Also, heuristics such as appending (part of) the conversation history to the current turn query are likely to lead to query drift~\cite{mitra_improving_1998}.
Recent work has modeled query resolution as a sequence generation task~\cite{raghu_statistical_2015,kumar_incomplete_2017,elgohary_can_2019}.
Another way of implicitly solving query resolution is by query modeling~\cite{guan2012effective,yang2015query,van2016lexical}, which has been studied and developed under the setup of session-based search~\cite{carterette2014overview,carterette2016evaluating}. 

\begin{table}[t]
	\centering
	\small
	\def\arraystretch{1}
	\caption{Excerpt from an example conversational dialog. Co-occurring terms in the conversation history and the relevant passage to the current turn (\#4) are shown in bold-face.
	}
	\begin{tabularx}{\linewidth}{l p{7cm}}
		\toprule
        \textbf{Turn} & \textbf{Query}
        \\\midrule
		1 & who formed \textbf{saosin}? \\
		2 & when was the \textbf{band} founded? \\
		3 & what was their \textbf{first} album?	\\ \hline
		4 & when was the album released? \\  		
		 & \emph{resolved:} when was saosin 's first album released? \\
		\midrule
		\multicolumn{2}{p{0.95\linewidth}}{\textit{Relevant passage to turn \#4}: The original lineup for \textbf{Saosin}, consisting of Burchell, Shekoski, Kennedy and Green, was formed in the summer of 2003. On June 17, the \textbf{band} released their \textbf{first} commercial production, the EP Translating the Name.}\\
        \bottomrule
	\end{tabularx}
	\label{tab:quac_paragraph1}
\end{table}
%
%

In this paper, we propose to model query resolution for conversational search as a binary term classification task: for each term in the previous turns of the conversation decide whether to add it to the current turn query or not.
We propose QuReTeC (\textbf{Qu}ery \textbf{Re}solution by \textbf{Te}rm \textbf{C}lassification), a query resolution model based on bidirectional transformers~\cite{vaswani_attention_2017} -- more specifically BERT~\cite{devlin_bert:_2019}.
The model encodes the conversation history and the current turn query and uses a term classification layer to predict a binary label for each term in the conversation history.
We integrate QuReTeC in a standard two-step cascade architecture that consists of an initial retrieval step and a reranking step.
This is done by using the set of terms predicted as relevant by QuReTeC as query expansion terms.
%

%

Training QuReTeC requires binary labels for each term in the conversation history.
One way to obtain such labels is to use human-curated gold standard query resolutions~\cite{elgohary_can_2019}.
However, these labels might be cumbersome to obtain in practice.
On the other hand, researchers and practitioners have been collecting general-purpose passage relevance labels, either by the means of human annotations or by the means of weak signals, e.g., clicks or mouse movements~\cite{joachims_optimizing_2002}.
We propose a distant supervision method to automatically generate training data, on the basis of such passage relevance labels. 
The key assumption is that passages that are relevant to the current turn share context with the conversation history that is missing from the current turn query.
Table~\ref{tab:quac_paragraph1} illustrates this assumption: the relevant passage to turn \#4 shares terms with the conversation history. 
Thus, we label the terms that co-occur in the relevant passages\footnote{A relevance passage contains not only the answer to the question but also context and supporting facts that allow the algorithm or the human to reach to this answer.} and the conversation history as relevant for the current turn.

Our main contributions can be summarized as follows:
\begin{enumerate}[leftmargin=*,nosep]
	\item We model the task of query resolution as a binary term classification task and propose to address it with a neural model based on bidirectional transformers, QuReTeC.
	\item We propose a distant supervision approach that can use general-purpose passage relevance data to substantially reduce the amount of human-curated data required to train QuReTeC.
	\item We experimentally show that when integrating the QuReTeC model in a multi-stage ranking architecture we significantly outperform baseline models. Also, we conduct extensive ablation studies and analyses to shed light into the workings of our query resolution model and its impact on retrieval performance.
\end{enumerate}
%
%

%

\section{Related work}

\paragraph{Conversational search}
Early studies on conversational search have focused on characterizing information seeking strategies and building interactive IR systems~\cite{oddy_information_1977,belkin1980anomalous,Croft:1987:IRN:35053.35054,belkin1995cases}.
\citet{Vtyurina:2017:ECS:3027063.3053175} investigated human behaviour in conversational systems through a user study and find that existing conversational assistants cannot be effectively used for conversational search with complex information needs.
\citet{Radlinski:2017:TFC:3020165.3020183} present a theoretical framework for conversational search, which highlights the need for multi-turn interactions.
\citet{cast2019} organize the Conversational Assistance Track (CAsT) at TREC 2019.
The goal of the track is to establish a concrete and standard collection of data with information needs to make systems directly comparable.
They release a mutli-turn passage retrieval dataset annotated by experts, which we use to compate our method to the baseline methods.
\paragraph{Query resolution}
Query resolution has been studied in the context of dialogue systems.
~\citet{raghu_statistical_2015} develop a pipeline model for query resolution in dialogues as text generation.
\citet{kumar_incomplete_2017} follow up on that work by using a sequence to sequence model combined with a retrieval model.
However, both these works rely on templates that are not available in our setting.
More related to our work, \citet{elgohary_can_2019} studied query resolution in the context of conversational QA over a single paragraph text.
They use a sequence to sequence model augmented with a copy and an attention mechanism and a coverage loss.
They annotate part of the QuAC dataset~\cite{choi_quac:_2018} with gold standard query resolutions on which they apply their model and obtain competitive performance.
In contrast to all the aforementioned works that model query resolution as text generation, we model query resolution as binary term classification in the conversation history.
\paragraph{Query modeling}
\label{sec:session-search}
Query modeling has been used in session search, where the task is to retrieve documents for a given query by utilizing previous queries and user interactions with the retrieval system~\cite{carterette2014overview}.
~\citet{guan2012effective} extract substrings from the current and previous turn queries to construct a new query for the current turn.
~\citet{yang2015query} propose a query change model that models both edits between consecutive queries and the ranked list returned by the previous turn query.
\citet{van2016lexical} compare the lexical matching session search approaches and find that naive methods based on term frequency weighing perform on par with specialized session search models.
The methods described above are informed by studies of how users reformulate their queries and why~\cite{DBLP:journals/ir/SloanYW15}, which, in principle, is different in nature from conversational search. 
For instance, in session search users tend to add query terms more than removing query terms, which is not the case in (spoken) conversational search.
Another form of query modeling is query expansion. 
Pseudo-relevance feedback is a query expansion technique that first retrieves a set of documents that are assumed to be relevant to the query, and then selects terms from the retrieved documents that are used to expand the query~\cite{Lavrenko:2001:RBL:383952.383972,abdul2004umass,nogueira_task-oriented_2017}.
Note that pseudo-relevance feedback is fundamentally different from query resolution: in order to revise the query, the former relies on the top-ranked documents, while the latter only relies on the conversation history.
\paragraph{Distant supervision}
Distant supervision can be used to obtain large amounts of noisy training data.
One of its most successful  applications is relation extraction, first proposed by~\citet{mintz_distant_2009}.
They take as input two entities and a relation between them, gather sentences where the two entities co-occur from a large text corpus, and treat those as positive examples for training a relation extraction system.
Beyond relation extraction, distant supervision has also been used to automatically generate noisy training data for other tasks such as named entity recognition~\cite{yang_distantly_2018}, sentiment classification~\cite{ritter_named_2011}, knowledge graph fact contextualization~\cite{voskarides-weakly-supervised-2018} and dialogue response generation~\cite{ren-2020-thinking}.
In our work, we follow the distant supervision paradigm to automatically generate training data for query resolution in conversational search by using query-passage relevance labels.
%
%

%

\section{Multi-turn Passage Retrieval Pipeline}
In this section we provide formal definitions and describe our multi-turn passage retrieval pipeline. 
Table~\ref{tab:notation} lists notation used in this paper.

 \begin{table}[t]
\caption{Notation used in the paper.}
\begin{tabularx}{\linewidth}{lp{6cm}}
\toprule
\bf Name  & \bf Description \\ 
\midrule
$terms(x)$  & set of terms in term sequence $x$\\    
$D$ & Passage collection \\ 
$q_i$ & Query at the current turn $i$\\
$q_{1:i-1}$ & Sequence of previous turn queries \\  
$q^*_i$ & Gold standard resolution of $q_i$\\
$E_{q_i}^*$   & Gold standard resolution terms for $q_i$, see Eq. \eqref{eq:gold-completion-terms} \\
$\hat{q}_i$ & Predicted resolution of $q_i$\\
$p^*_{q_i}$ & A relevant passage for $q_i$\\
\bottomrule 
\end{tabularx}
\label{tab:notation}
\end{table}

\subsection{Definitions}
\label{sec:definitions}

\paragraph{Multi-turn passage ranking} 
Let $[ q_1, \ldots, q_{i-1}, q_i ]$ be a sequence of conversational queries that share a common topic $T$.
Let $q_i$ be the current turn query and $q_{1:i-1}$ be the conversation history.
Given $q_i$ and $q_{1:i-1}$, the task is to retrieve a ranked list of passages $L$ from a passage collection $D$ that satisfy  the user's information need.\footnote{We follow the TREC CAsT setup and only take into account $q_{1:i-1}$ but not the passages retrieved for $q_{1:i-1}$.}

In the multi-turn passage ranking task, the current turn query $q_i$ is often underspecified due to phenomena such as zero anaphora, topic change, and topic return.
Thus, context from the conversation history $q_{1:i-1}$ must be taken into account to arrive at a better expression of the current turn query $q_i$.
This challenge can be addressed by query resolution.

\paragraph{Query resolution} 
Given the conversation history $q_{1:i-1}$ and the current turn query $q_i$, output a query $\hat{q}_i$ that includes both the existing information in $q_i$ and the missing context of $q_i$ that exists in the conversation history $q_{1:i-1}$.

\begin{figure}[t]
  \centering
   \includegraphics[scale=0.38]{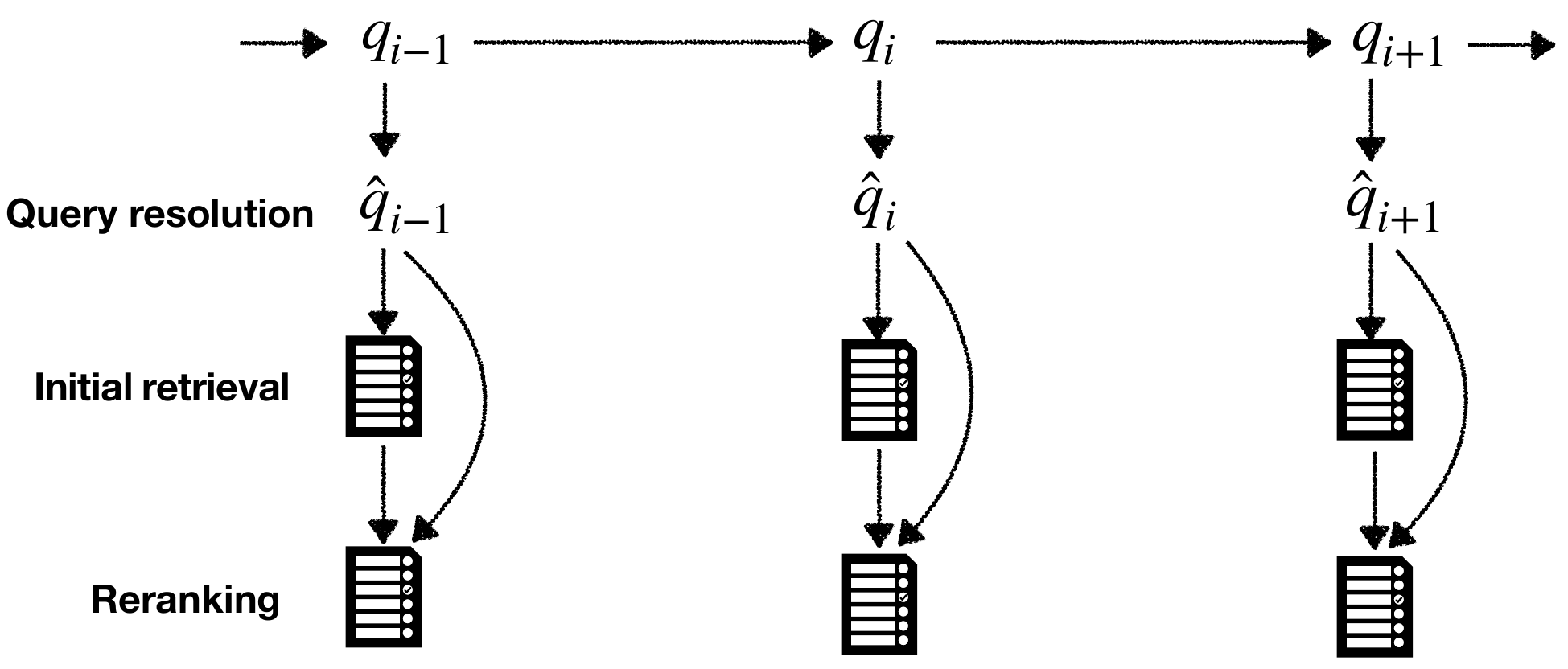}
   \caption{Illustration of our multi-turn passage retrieval pipeline for three turns.}
\label{fig:pipeline}
\end{figure}

\subsection{Multi-turn passage retrieval pipeline}

Figure~\ref{fig:pipeline} illustrates our multi-turn passage retrieval pipeline.
We use a two-step cascade ranking architecture~\cite{wang2011cascade}, which we augment with a query resolution module (Section~\ref{sec:query-resolution}). First, the unsupervised initial retrieval step outputs the initial ranked list $L_1$ (Section~\ref{sec:initial-retrieval}). Second, the re-ranking step outputs the final ranked list $L$ (Section~\ref{sec:reranking}).
Below we describe the two steps of the cascade ranking architecture.

\subsubsection{Initial retrieval step}
\label{sec:initial-retrieval}

In this step we obtain the initial ranked list $L_1$ by scoring each passage $p$ in the passage collection $D$ with respect to the resolved query $\hat{q}_i$ using a lexical matching ranking function $f_1$.
We use query likelihood (QL) with Dirichlet smoothing~\cite{zhai2004study} as $f_1$, since it 
 outperformed other ranking functions such as BM25 in preliminary experiments over the TREC CAsT dataset.

\subsubsection{Reranking step}
\label{sec:reranking}

In this step, we re-rank the list $L_1$ by scoring each passage $p \in L_1$ with a ranking function $f_2$ to obtain the final ranked list $L$.
To construct $f_2$, we use rank fusion and combine the scores obtained by $f_1$ (used in initial retrieval step) and a supervised neural ranker $f_n$.
Next, we describe the neural ranker $f_n$.

\paragraph{Supervised neural ranker}
We use BERT~\cite{devlin_bert:_2019} as the neural ranker $f_n$, as it has been shown to achieve state-of-the-art performance in ad-hoc retrieval~\cite{macavaney_cedr:_2019,qiao_understanding_2019,yang2019simple}.
Also, BERT has been shown to prefer semantic matches~\cite{qiao_understanding_2019}, and thereby can be complementary to $f_1$, which is a lexical matching method.
As is standard when using BERT for pairs of sequences, the input to the model is formatted as [ \texttt{<CLS>},  $\hat{q}_i$  \texttt{<SEP>}, $p$], where $\texttt{<CLS>}$ is a special token, $\hat{q}_i$ is the resolved current turn query, $p$ is the passage.
We add a dropout layer and a linear layer $l_a$ on top of the representation of the \texttt{<CLS>} token in the last layer, followed by a $\tanh$ function to obtain $f_n$~\cite{macavaney_cedr:_2019}.
We score each passage $p \in L_1$ using $f_n$ to obtain $L_n$  .
We fine-tune the pretrained BERT model using pairwise ranking loss on a large-scale single-turn passage ranking dataset~\cite{yang2019simple}.
During training we sample as many negative as positive passages per query.

\paragraph{Rank fusion}
We design $f_2$ such that it combines lexical matching and semantic matching~\cite{onal-neural-2018}.
We use Reciprocal Rank Fusion (RRF)~\cite{cormack_reciprocal_2009} to combine the score obtained by the lexical matching ranking function $f_1$, and the semantic matching supervised neural ranker $f_n$.
We choose RRF because of its effectiveness in combining individual rankers in ad-hoc retrieval and because of its simplicity (it has only one hyper-parameter).
We define $f_2$ as the RRF of $L_1$ and $L_n$~\cite{cormack_reciprocal_2009}:  
\begin{align}
	\label{eq:recip_rank_fusion}
	f_2(p) = \sum_{L' \in \{L_1, L_n\}} \frac{1}{k + rank(p, L')},
\end{align}
where $rank(p, L')$ is the rank of passage $p$ in a ranked list $L'$, and $k$ is a hyperparameter.\footnote{We set $k=60$ and do not tune it.} We score each passage $p$ in the initial ranked list $L_1$ with $f_2$ to obtain the final ranked list $L$. 

Since developing specialized re-rankers for the task at hand is not the focus of this paper, we leave more sophisticated methods for choosing the neural ranker $f_n$ and for combining multiple rankers as future work. 
In the next section, we describe our query resolution model, QuReTeC, which is the focus of this paper.

%

\section{Query Resolution}
\label{sec:query-resolution}

In this section we first describe how we model query resolution as term classification (Section~\ref{sec:query-resolution-as-tc}), then present our query resolution model, QuReTeC, (Section~\ref{sec:predicting-relevant-terms}), and finally describe how we generate distant supervision labels for the model (Section~\ref{sec:distant-supervision-for-query-resolution}).

\subsection{Query resolution as term classification in the conversation history}
\label{sec:query-resolution-as-tc}
Previous work has modeled query resolution as a sequence to sequence task~\cite{kumar_incomplete_2017,elgohary_can_2019}, 
where the source sequence is $q_{1:i}$ and the target sequence is $q^*_i$, where $q^*_i$ is a gold standard resolution of the current turn query $q_i$.
For instance, the gold standard resolution of turn \#4 in Table~\ref{tab:quac_paragraph1} is: ``When was Saosin's first album released?''

However, since (i) the initial retrieval step of our pipeline (Section~\ref{sec:initial-retrieval}) is a term-based model that treats queries as bag of words, and (ii) the supervised neural ranker we use in the re-ranking step (Section~\ref{sec:reranking}) is robust to queries that are not well-formed natural language texts~\cite{yang2019simple}, our query resolution model does not necessarily need to output a well-formed natural language query but rather a set of terms to expand the query.
Besides, sequence to sequence based models generally need a massive amount of data for training in order to get reasonable performance due to their generation objective~\cite{fadaee_data_2017}.
Therefore, we model query resolution as a term classification task: given the conversation history $q_{1:i-1}$ and the current turn query $q_i$, output a binary label (relevant or non-relevant) for each term in $q_{1:i-1}$.
Terms in the conversation history $q_{1:i-1}$ that are tagged as relevant are appended to the current turn query $q_i$ to form the predicted current turn query resolution $\hat{q}_i$.

We define the set of relevant resolution terms $E^*(q_i)$ as:
\begin{align}
	\label{eq:gold-completion-terms}
	 E_{q_i}^* = terms(q^*_i) \cap terms(q_{1:i-1}) \setminus terms(q_i),
\end{align}
where $q^*_i$ is a gold standard resolution of the current turn query $q_i$.
Under this formulation, the set of relevant terms $E_{q_i}^*$ represents the missing context from the conversation history $q_{1:i-1}$.
For instance, the set of gold standard resolution terms $E_{q_i}^*$ for turn \#4 in Table~\ref{tab:quac_paragraph1} is $\{  \text{Saosin}, \text{first} \}$.
Note that $E_{q_i}^*$ can be empty if $q_i = q^*_i$, i.e., the current turn query does not need to be resolved, or if $terms(q^*_i) \cap terms(q_{1:i-1})$ is empty. In our experiments $terms(q^*_i) \cap terms(q_{1:i-1}) \approx terms(q^*_i)$, and therefore almost all the gold standard resolution terms can be found in the conversation history.

\subsection{Query resolution model}
\label{sec:predicting-relevant-terms}
\begin{figure*}[t]
    \centering
    \begin{subfigure}[t]{0.2\textwidth}
        \centering
        \includegraphics[scale=0.4]{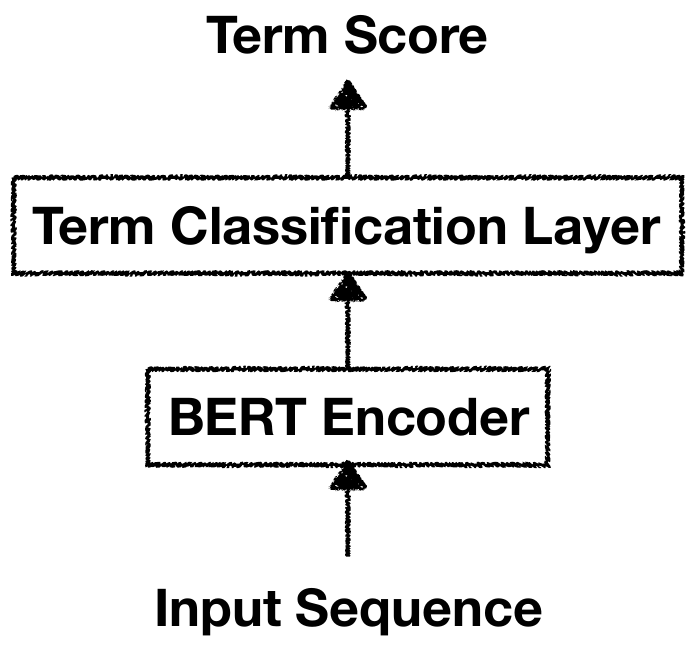}
    	\caption{QuReTeC model architecture.}
	\label{fig:model_architecture}
    \end{subfigure}   
    ~
    \begin{subfigure}[t]{0.8\textwidth}
        \centering
        \includegraphics[scale=0.35]{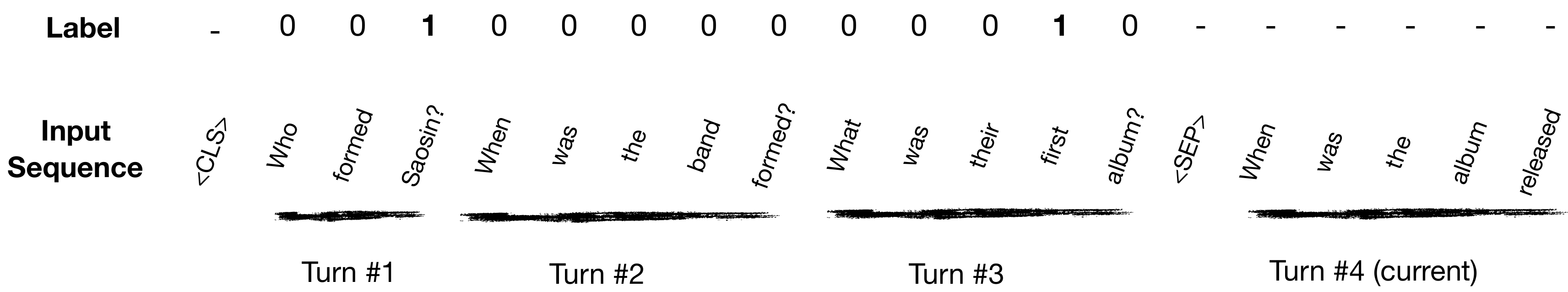}
    	\caption{Example input sequence and gold standard term labels (1: relevant, 0: non-relevant) for QuReTeC.}
	\label{fig:model_input_output}
    \end{subfigure}
    \caption{}
    \label{fig:model_architecture_and_input_output}
\end{figure*}
In this section, we describe our query resolution model, QuReTeC.
Figure~\ref{fig:model_architecture} shows the model architecture of QuReTeC.
Each term in the input sequence is first encoded using bidirectional transformers~\cite{vaswani_attention_2017} -- more specifically BERT~\cite{devlin_bert:_2019}. 
Then, a term classification layer takes each encoded term as input and outputs a score for each term. 
We use BERT as the encoder since it has been successfully applied in tasks similar to ours, such as named entity recognition and coreference resolution~\cite{devlin_bert:_2019,joshi_bert_2019,lin2019open}.
Next we describe the main parts of QuReTeC in detail, i.e., input sequence, BERT encoder and Term classification layer.

\begin{enumerate}[leftmargin=*,nosep]
\item \textit{Input sequence.}
The input sequence consists of all the terms in the queries of the previous turns $q_{1:i-1}$ and the current turn $q_i$.
It is formed as: [\texttt{<CLS>}, $terms(q_1)$, \ldots, $terms(q_{i-1})$,  \texttt{<SEP>}, $terms(q_i)$], where \texttt{<CLS>} and \texttt{<SEP>} are special tokens.
We add a special separator token \texttt{<SEP>} between the previous turn $q_{i-1}$ and the current turn $q_i$ in order to inform the model where the current turn begins.
Figure~\ref{fig:model_input_output} shows an example input sequence and the gold standard term labels.

\item \textit{BERT encoder.}
BERT first represents the input terms with WordPiece embeddings using a 30K vocabulary.
After applying multiple transformer blocks, BERT outputs an encoding for each term.
We refer the interested reader to the original paper for a detailed description of BERT~\cite{devlin_bert:_2019}.

\item \textit{Term classification layer.}
The term classification layer is applied on top of the representation of the first sub-token of each term~\cite{devlin_bert:_2019}.
It consists of a dropout layer, a linear layer and a sigmoid function and outputs a scalar for each term.
We mask out the output of \texttt{<CLS>}
and the current turn terms, since we are not interested in predicting a label for those (see Equation~\eqref{eq:gold-completion-terms} for the definition and Figure~\ref{fig:model_input_output} for an example).
\end{enumerate}

In order to train QuReTeC we need a dataset containing gold standard resolution terms $E_{q_i}^*$ for each $q_i$. 
The terms in $E_{q_i}^*$ are labeled as relevant and the rest of the terms ($terms(q_{1:i-1}) \setminus E_{q_i}^*$) as non-relevant.
Assuming there exists a gold standard resolution $q^*_i$ for each $q_i$, we can derive $E_{q_i}^*$  using Equation~\eqref{eq:gold-completion-terms}.
We use standard binary cross entropy as the loss function. 

\subsection{Generating distant supervision for query resolution}
\label{sec:distant-supervision-for-query-resolution}
Recall that the gold standard resolution $q^*_i$ includes the information in $q_i$ and the missing context of $q_i$ that exists in the conversation history $q_{1:i-1}$.
As described above, we can train QuReTeC if we have a gold standard resolution $q^*_i$ for each $q_i$.
Obtaining such special-purpose gold standard resolutions is cumbersome compared to almost readily available general-purpose passage relevance labels for $q_i$.
We propose a distant supervision method to generate labels to train QuReTeC.
Specifically, we simply replace $q^*_i$ with a relevant passage $p^*_{q_i}$ in Equation~\eqref{eq:gold-completion-terms} to extract the set of relevant resolution terms $E_{q_i}^*$.
Table~\ref{tab:quac_paragraph1} illustrates this idea with an example dialogue and the relevant passage to the current turn query.
The gold standard resolution terms extracted with this distant supervision procedure for this example are $\{\text{Saosin}, \text{first}, \text{band}\}$.

Intuitively, the above procedure is noisy and can result in adding terms to $E_{q_i}^*$ that are non-relevant, or adding too few relevant terms to $E_{q_i}^*$.
Nevertheless, we experimentally show in Section~\ref{sec:distant-supervision-query-resolution} that this distant supervision signal can be used to substantially reduce the number of human-curated gold standard resolutions required for training QuReTeC.

The distant supervision method we describe here makes QuReTeC more generally applicable than other supervised methods such as the method in~\citet{elgohary_can_2019} that can only be trained with gold standard query resolutions.
This is because, apart from manual annotation, query-passage relevance labels can be potentially obtained at scale by using click logs~\cite{joachims_optimizing_2002}, or weak supervision~\cite{dehghani_neural_2017}.

%
\section{Experimental setup}

\subsection{Research questions}
We aim to answer the following research questions:
\begin{enumerate}[label=(RQ\arabic*),leftmargin=*,nosep]

\item How does the QuReTeC model perform compared to other state-of-the-art methods?

\item Can we use distant supervision to reduce the amount of human-curated training data required to train QuReTeC?

\item How does QuReTeC's performance vary depending on the turn of the conversation?

\end{enumerate}
For all the research questions listed above we measure performance in both an intrinsic and an extrinsic sense.
\emph{Intrinsic} evaluation measures query resolution performance on term classification.  
\emph{Extrinsic} evaluation measures retrieval performance at both the initial retrieval and the reranking steps. 

\subsection{Datasets}

\begin{table*}[t]
\caption{TREC CAsT 2019 multi-turn passage retrieval dataset statistics.}
\begin{tabularx}{0.85\textwidth}{l c c r@{ $\pm$ }l r@{ $\pm$ }l r@{ $\pm$ }l r@{ $\pm$ }l }
\toprule
Split & \#Topics & \#Queries & 
 \multicolumn{2}{c}{\tabincell{l}{\#Labelled passages\\ per topic}}  & 
 \multicolumn{2}{c}{\tabincell{l}{\#Relevant passages \\ per topic}}   &  
 \multicolumn{2}{c}{\tabincell{l}{\#Labelled passages\\ per query}}  &  
 \multicolumn{2}{c}{\tabincell{l}{ \#Relevant passages\\ per query}} \\ 
\midrule
Test  & 20 & 173  &  1,467.50 & 252.86 & 406.00 & 190.18 & 169.65 & 36.69 & 46.94 & 31.53  \\
\bottomrule
\end{tabularx}
\label{tbl:passage-retrieval-data}
\end{table*}

\begin{table}[t]
\caption{Query resolution datasets statistics. In the Split column, we indicate the where the positive term labels originate from: either gold (gold standard resolutions) or distant (Section~\ref{sec:distant-supervision-for-query-resolution}).}
\setlength{\tabcolsep}{4pt}
\begin{tabularx}{\linewidth}{ l l r r@{ $\pm$ }l r@{ $\pm$ }l}
\toprule
Dataset   & Split & \#Queries & \multicolumn{4}{c}{\#Terms (per query)}       \\ 
\cmidrule(l){4-7}
 & &  & \multicolumn{2}{c}{Total} & \multicolumn{2}{c}{Positive}   \\
\midrule
QuAC & Train (gold)   &   20,181 & 97.96 & 61.02 & 4.56 & 3.88     \\
& Train (distant)   &   31,538 & 99.78 & 62.36 & 6.90 & 5.59   \\
& Dev (gold)   &   2,196 & 95.49 & 58.79 & 4.49 & 3.90      \\
& Test (gold)   &    3,373 & 96.96 & 59.24 & 4.30 & 3.86      \\ 
\midrule
CAsT & Test  (gold)  & 153 & 39.97 & 17.97 & 1.89 & 1.62    \\
\bottomrule 
\end{tabularx}
\label{tbl:question-resolution-datasets}
\vspace{-4mm}
\end{table}

\subsubsection{Extrinsic evaluation -- retrieval}
The TREC CAsT dataset is a multi-turn passage retrieval dataset~\cite{cast2019}.
It is the only such dataset that is publicly available.
Each topic consists of a sequence of queries.
The topics are open-domain and diverse in terms of their information need.
The topics are curated manually to reflect information seeking conversational structure patterns.
Later turn queries in a topic depend only on the previous turn queries, and not on the returned passages of the previous turns, which is a limitation of this dataset.
Nonetheless, the dataset is sufficiently challenging for comparing automatic systems, as we will show in Section~\ref{sec:query-resolution-reranking}.
Table~\ref{tbl:passage-retrieval-data} shows statistics of the dataset.
The original dataset consists of 30 training and 50 evaluation topics.
20 of 50 topics in the evaluation set were annotated  for relevance by NIST assessors on a 5-point relevance scale.
We use this set as the TREC CAsT test set.
The organizers also provided a small set of judgements for the training set, however we do not use it in our pipeline.
%
%
The passage collection is the union of two passage corpora, the MS MARCO~\cite{MSMARCO} (Bing), and the TREC CAR~\cite{treccar} (Wikipedia passages).\footnote{The Washington Post collection was also part of the original collection but it was excluded from the official TREC evaluation process and therefore we do not use it.}

\subsubsection{Intrinsic evaluation -- query resolution}
\label{sec:datasets-intrinsic}
The original QuAC dataset~\cite{choi_quac:_2018}  contains dialogues on a single Wikipedia article section regarding people (e.g., early life of a singer).
Each dialogue contains up to 12 questions and their corresponding answer spans in the section.
It was constructed by asking two crowdworkers (a student and a teacher) to perform an interactive dialogue about a specific topic.
\citet{elgohary_can_2019} crowdsourced question resolutions for a subset of the original QuAC dataset~\cite{choi_quac:_2018}.
All the questions in the \emph{dev} and \emph{test} splits of~\cite{elgohary_can_2019} have gold standard resolutions. 
We use the \emph{dev} split for early stopping when training QuReTeC and evaluate on the \emph{test} set.
When training with gold supervision (gold standard query resolutions), we use the \emph{train} split from~\citep{elgohary_can_2019}, which is a subset of the train split of~\cite{choi_quac:_2018}; all the questions therein have gold standard resolutions.
Since QuAC is not a passage retrieval collection, in order to obtain distant supervision labels (Section~\ref{sec:distant-supervision-for-query-resolution}), we use a window of 50 characters around the answer span to extract passage-length texts, and we treat the extracted passage as the relevant passage.
When training with distant labels, we use the part of the \emph{train} split of~\cite{choi_quac:_2018} that does not have gold standard resolutions.

The TREC CAsT dataset~\cite{cast2019} also contains gold standard query resolutions for its test set. However, it is too small to train a supervised query resolution model, and we only use it as a complementary \emph{test} set. 

The two query resolution datasets described above have three main differences.
First, the conversations in QuAC are centered around a single Wikipedia article section about people whereas the conversations in CAsT are centered around an arbitrary topic. 
Second, the answers of the QuAC questions are spans in the Wiki\-pedia section whereas the CAsT queries have relevant passages that originate from different Web resources besides Wikipedia. 
Third, later turns in QuAC do depend on the answers in previous turns, while in CAsT they do not (Section~\ref{sec:definitions}).
Interestingly, in Section~\ref{sec:results-query-res-multi} we demonstrate that despite these differences, training QuReTeC on QuAC generalizes well to the CAsT dataset.

Table~\ref{tbl:question-resolution-datasets} provides statistics for the two datasets.\footnote{
Note that the first turn in each topic does not need query resolution because there is no conversation history at that point and thus the query resolution CAsT test has 20 (the number of topics) fewer queries than in Table~\ref{tbl:passage-retrieval-data}.}
First, we observe that the QuAC dataset is much larger than CAsT.
Also, QuAC has a larger number of terms on average than CAsT (\textasciitilde97 vs \textasciitilde40) and a larger negative-positive ratio (\textasciitilde20:1 vs \textasciitilde40:1).
This is because in QuAC the answers to the previous turns are included in the conversation history whereas in CAsT they are not. For this reason, we expect query resolution on QuAC to be more challenging than on CAsT.

\subsection{Evaluation metrics}
\subsubsection{Extrinsic evaluation -- retrieval}
We report NDCG@3 (the official TREC CAsT evaluation metric), Recall, MAP, and MRR at rank 1000.
We also provide performance metrics averaged per turn to show how retrieval performance varies across turns.

We report on statistical significance with a paired two-tailed t-test. We depict a significant increase for $p<0.01$ as $^\blacktriangle$.

\subsubsection{Intrinsic evaluation -- query resolution}
We report on Micro-Precision (P), Micro-Recall (R) and Micro-F1 (F1), i.e., metrics calculated per query and then averaged across all turns and topics. We ignore queries that are the first turn of the conversation when calculating the mean, since we do not predict term labels for those.

\subsection{Baselines}
We perform intrinsic and extrinsic evaluation by comparing against a number of query resolution baselines.
Next, we provide a detailed description of each baseline:
\begin{itemize}[leftmargin=*,nosep]
	\item \textbf{Original} This method uses the original form of the query. We explore different variations for constructing $\hat{q}_i$: \begin{enumerate*} \item current turn only (cur), \item current turn expanded by the previous turn (cur+prev), \item current turn expanded by the first turn (cur+first), and \item all turns.\end{enumerate*}

	\item \textbf{RM3~\cite{abdul2004umass}} A state-of-the-art unsupervised pseudo-relevance feedback model.\footnote{Note that given the very small size of the TREC CAsT training set we do not compare to more sophisticated yet data-hungry pseudo-relevance feedback models such as~\cite{nogueira_task-oriented_2017}.} RM3 first performs retrieval and treats the top-$n$ ranked passages as relevant. Then, it estimates a query language model based on the top-$n$ results, and finally adds the top-$k$ terms to the original query. As with Original, we report on different variations for constructing the query: cur, cur+prev, cur+first and all turns. In order to apply RM3 for query resolution we append the top-$k$ terms to the original query $q_i$ to obtain $\hat{q}_i$.

	\item \textbf{NeuralCoref}\footnote{\url{https://medium.com/huggingface/state-of-the-art-neural-coreference-resolution-for-chatbots-3302365dcf30}}  A coreference resolution method designed for chatbots. It uses a rule-based system for mention detection and a feed-forward neural network that predicts coreference scores. We perform coreference resolution on the conversation history $q_{1:i-1}$ and the current turn query $q_i$. The output $\hat{q}_i$ consists of $q_i$ and the predicted terms in $q_{1:i-1}$ where terms in $q_i$ refer to.

	\item \textbf{BiLSTM-copy~\cite{elgohary_can_2019}} A neural sequence to sequence model for query resolution. It uses a BiLSTM encoder and decoder augmented with attention and copy mechanisms and also a coverage loss~\cite{see_get_2017}. It initializes the input embeddings with pretrained GloVe embeddings.\footnote{\url{https://nlp.stanford.edu/projects/glove/}} Given $q_{1:i-1}$ and  $q_i$, it outputs $\hat{q}_i$. It was optimized on the QuAC gold standard resolutions.

\end{itemize}

\subsubsection{Intrinsic evaluation -- query resolution}
In order to perform intrinsic evaluation on the aforementioned baselines, we take the query resolution they output ($\hat{q}_i$) and apply Equation \eqref{eq:gold-completion-terms} by replacing $q^*_i$ with $\hat{q}_i$ to obtain the set of predicted resolution terms.

\subsubsection{Extrinsic evaluation -- initial retrieval}
Here, apart from the aforementioned baselines, we also use the following baselines:
\begin{itemize}[leftmargin=*,nosep]
	\item \textbf{Nugget}~\cite{guan2012effective}. Extracts substrings from the current and previous turn queries to build a new query for the current turn.\footnote{We use the nugget version that does not depend on anchors text since they are not available in our setting.}
	\item \textbf{QCM}~\cite{yang2015query}. Models the edits between consecutive queries and the results list returned by the previous turn query to construct a new query for the current turn. 
	\item \textbf{Oracle} Performs initial retrieval using the gold standard resolution query. Released by the TREC CAsT organizers.
\end{itemize}

\subsubsection{Extrinsic evaluation -- reranking}
Since developing specialized rerankers for multi-turn passage retrieval is not the focus of this paper, we evaluate the reranking step using ablation studies.
For reference, we also report on the performance of the top-ranked TREC CAsT 2019 systems~\cite{cast2019}:
\begin{itemize}[leftmargin=*,nosep]
	\item \textbf{TREC-top-auto} Uses an automatic system for query resolution and BERT-large for reranking.
	\item \textbf{TREC-top-manual} Uses the gold standard query resolution and BERT-large for reranking. 
\end{itemize}

\subsection{Implementation \& hyperparameters}
\textbf{Multi-turn passage retrieval}
We index the TREC CAsT collections using Anserini with stopword removal and stemming.\footnote{\url{https://github.com/castorini/anserini}}
In the initial retrieval step (section~\ref{sec:initial-retrieval}) we retrieve the top 1000 passages using QL with Dirichlet smoothing (we set $\mu = 2500$).
We use the default value for the fusion parameter $k=60$~\cite{cormack_reciprocal_2009}  in Eq.~\eqref{eq:recip_rank_fusion}.
In the reranking step (section~\ref{sec:reranking}) we use a PyTorch implementation of BERT for retrieval~\cite{macavaney_cedr:_2019}.
We use the \texttt{bert-base-uncased} pretrained BERT model.
We fine-tune the BERT reranker with MSMARCO passage ranking dataset~\cite{bajaj_ms_2018}. We train on 100K randomly sampled training triples from its training set and evaluate on 100 randomly sampled queries of its development set.
We use the Adam optimizer with a learning rate of $0.001$ except for the BERT layers for which we use a learning rate of $3\mathrm{e}{-6}$. We apply dropout with a probability of $0.2$ on the output linear layer.
We apply early stopping on the development set with a patience of 2 epochs based on MRR.
\textbf{Query resolution}
We use the \texttt{bert-large-uncased} model.
We implement QuReTeC on top of HuggingFace's PyTorch implementation of BERT.\footnote{\url{https://github.com/huggingface/transformers}} 
We use the Adam optimizer and tune the learning rate in the range $\{ 2\mathrm{e}{-5}$, $3\mathrm{e}{-5}$, $3\mathrm{e}{-6} \}$.
We use a batch size of 4 and do gradient clipping with the value of $1$.
We apply dropout on the term classification layer and the BERT layers in the range $\{0.1, 0.2, 0.3, 0.4\}$.
We optimize for F1 on the QuAC dev (gold) set.

\textbf{Baselines}
\label{sec:exp-setup-baselines}
For RM3, we tune the following parameters: $n \in  \{3, 5, 10, 20, 30\}$ and $k \in \{5, 10\}$ and set the original query weight to the default value of $0.8$.
For Nugget, we set $k_{snippet}=10$ and tune $\theta \in \{0.95, 0.97,0.99\}$. 
For QCM, we tune $\alpha \in \{ 1.0, 2.2, 3.0\} $, $\beta \in \{1.6, 1.8, 2.0 \}$, $\epsilon \in \{0.06, 0.07, 0.08\}$ and $\delta \in \{0.2, 0.4, 0.6\}$.
For both Nugget and QCM we use \citet{van2016lexical}'s implementation.  For fair comparison, we retrieve over the whole collection rather than just reranking the top-1000 results.
The aformentioned methods are tuned on the small annotated training set of TREC CAsT.
For query resolution, we tune the greedyness parameter of NeuralCoref in the range $\{0.5, 0.75\}$.
We use the model of BiLSTM-copy released by ~\cite{elgohary_can_2019}, as it was optimized specifically for QuAC with gold standard resolutions.
\textbf{Preprocessing}
We apply lowercase, lemmatization and stopword removal to $q^*_i$, $q_{1:i-1}$ and $q_i$ using Spacy\footnote{\url{http://spacy.io/}} before calculating term overlap in Equation~\ref{eq:gold-completion-terms}.
%

%
\section{Results \& discussion}
\label{sec:results}
In this section we present and discuss our experimental results.
\subsection{Query resolution for multi-turn retrieval}
\label{sec:results-query-res-multi}
In this subsection we answer (RQ1): we study how QuReTeC performs compared to other state-of-the-art methods when evaluated on term classification (Section~\ref{sec:query-resolution-classification}), when incorporated in the initial retrieval step (Section~\ref{sec:query-resolution-initial-retrieval}) and in the reranking step (Section~\ref{sec:query-resolution-reranking}).

\subsubsection{Intrinsic evaluation}
\label{sec:query-resolution-classification}
\begin{table}[t]
\centering
\caption{Intrinsic evaluation for query resolution on the QuAC test set. Cur, prev, first and all refer to using the current, previous, first or all turns respectively.}
\label{tab:query-resolution-quac}
\begin{tabular}{llll}
\hline
\textbf{Method}  & \textbf{P} & \textbf{R} & \textbf{F1} \\ \hline
Original (cur+prev) & 22.3 & 46.4 & 30.1 \\
 Original (cur+first) & 41.1 & 49.5 & 44.9 \\
 Original (all) & 12.3 & \bf 100.0 & 21.9 \\  \hline
NeuralCoref  & 65.5 & 30.0 & 41.2 \\
BiLSTM-copy  &  67.0 & 53.2 & 59.3 \\ \hline
 
QuReTeC & \bf 71.5 & 66.1 & \bf 68.7 \\ \hline
\end{tabular}

\end{table}
\begin{table}[t]
\centering
\caption{Intrinsic evaluation for query resolution on the TREC CAsT test set. Cur, prev, first and all refer to using the current, previous, first, or all turns respectively.}
\label{tab:query-resolution-cast}
\begin{tabular}{llll}
\hline
\textbf{Method}  & \textbf{P} & \textbf{R} & \textbf{F1} \\ \hline
Original (cur+prev) & 32.5 & 43.9 & 37.4 \\
Original (cur+first) & 43.0 & 74.0 & 54.4 \\
Original (all) & 18.6 & \bf 100.0 & 31.4 \\ \hline
RM3 (cur) & 35.8 & 8.3 & 13.5 \\
 RM3 (cur+prev) & 34.6 & 32.5 & 33.5 \\
 RM3 (cur+first) & 40.9 & 32.9 & 36.5 \\
 RM3 (all) & 41.5 & 38.8 & 40.1 \\  \hline
NeuralCoref & \bf 83.0 & 28.7 & 42.7 \\
BiLSTM-copy  & 51.5 & 36.0 & 42.4 \\ \hline
QuReTeC  & 77.2 & 79.9 & \bf 78.5 \\ \hline
\end{tabular}
\end{table}
In this experiment we evaluate query resolution as a term classification task.\footnote{Note that the performance of Original (cur) is zero by definition when using the current turn only  (see Eq.~\ref{eq:gold-completion-terms}). Thus, we do not include it in Tables~\ref{tab:query-resolution-quac} and ~\ref{tab:query-resolution-cast}. Also, RM3 is not applicable in Table~\ref{tab:query-resolution-quac} since QuAC is not a retrieval dataset.}
Table~\ref{tab:query-resolution-quac} shows the query resolution results on the QuAC dataset.
We observe that QuReTeC outperforms all the variations of Original and the NeuralCoref by a large margin in terms of F1, precision and recall -- except for Original (all) that has perfect recall but at the cost of very poor precision.
Also, QuReTeC substantially outperforms BiLSTM-copy on all metrics. Note that BiLSTM-copy was optimized on the same training set as QuReTeC (see Section~\ref{sec:exp-setup-baselines}).
This shows that QuReTeC is more effective in finding missing contextual information from previous turns.
Table~\ref{tab:query-resolution-cast} shows the query resolution results on the CAsT dataset. 
Generally, we observe similar patterns in terms of overall performance as in Table~\ref{tab:query-resolution-quac}.
Interestingly, we observe that QuReTeC generalizes very well to the CAsT dataset (even though it was only trained on QuAC) and outperforms all the baselines in terms of F1 by a large margin.
In contrast, BiLSTM-copy fails to generalize and performs worse than Original (cur+first) in terms of F1.
NeuralCoref has higher precision but much lower recall compared to QuReTeC.
Finally, RM3 has relatively poor query resolution performance. This indicates that pseudo-relevance feedback is not suitable for the task of query resolution.

\subsubsection{Query resolution for initial retrieval}
\label{sec:query-resolution-initial-retrieval}
\begin{table}[t]
\centering
\caption{Initial retrieval performance on the TREC CAsT test set for different query resolution methods. The retrieval model is fixed (same as in Section~\ref{sec:initial-retrieval}). 
Significance is tested against RM3 (cur+first) since it has the best NDCG@3 among the baselines.}
\label{tab:initial-retrieval}
\begin{tabular}{lllll}
\hline
\textbf{Method} & \textbf{Recall} & \textbf{MAP} & \textbf{MRR} & \textbf{NDCG@3} \\ \hline
Original (cur) & 0.438 & 0.129 & 0.310 & 0.155 \\
 Original (cur+prev) & 0.572 & 0.181 & 0.475 & 0.235 \\
Original (cur+first) & 0.655 & 0.214 & 0.561 & 0.282 \\
 Original (all) & 0.694 & 0.190 & 0.552 & 0.256 \\ \hline
RM3  (cur) & 0.440 & 0.140 & 0.320 & 0.158 \\
 RM3 (cur+prev) & 0.575 & 0.200 & 0.482 & 0.254 \\
 RM3 (cur+first) & 0.656 & 0.225 & 0.551 & 0.300 \\
 RM3 (all) & 0.666 & 0.195 & 0.544 & 0.266 \\ \hline
Nugget & 0.426 & 0.101 & 0.334 & 0.145 \\
QCM  & 0.392 & 0.091 & 0.317 & 0.127 \\ \hline
NeuralCoref & 0.565 & 0.176 & 0.423 & 0.212 \\
BiLSTM-copy & 0.552 & 0.171 & 0.403 & 0.205 \\ \hline
QuReTeC  & \textbf{0.754}$^\blacktriangle$ & \textbf{0.272}$^\blacktriangle$ & \textbf{0.637}$^\blacktriangle$ & \textbf{0.341}$^\blacktriangle$ \\ \hline
Oracle & 0.785 & 0.309 & 0.660 & 0.361  \\ \hline

\end{tabular}
\vspace{-4mm}

\end{table}
In this experiment, we evaluate query resolution when incorporated in the initial retrieval step (Section~\ref{sec:initial-retrieval}).
We compare QuReTeC to the baseline methods in terms of initial retrieval performance.
Table~\ref{tab:initial-retrieval} shows the results.
First, we observe that QuReTeC outperforms all the baselines by a large margin on all metrics.  
Also, interestingly, QuReTeC achieves performance close to the one achieved by the Oracle performance (gold standard resolutions).
Note that there is still plenty of room for improvement even when using Oracle, which indicates that exploring other ranking functions for initial retrieval is a promising direction for future work.
QuReTeC outperforms all Original and RM3 variations, which perform similarly.
The session search methods (Nugget and QCM) perform poorly even compared to the Original variations, which indicates that session search is different in nature than conversational search. 
BiLSTM-copy performs poorly compared to QuReTeC but also compared to the Original variations, which means that it does not generalize well to CAsT.

\subsubsection{Query resolution for reranking}
\label{sec:query-resolution-reranking}
\begin{table}[t]
\centering
\caption{Reranking performance on the TREC CAsT test set. 
All the methods in the first group use QuReTeC for query resolution. 
Significance is tested against BERT-base.}
\label{tab:reranking}
\begin{tabularx}{\linewidth}{llll}
\hline
\textbf{Method} & \textbf{MAP} & \textbf{MRR} & \textbf{NDCG@3} \\ \hline
Initial & 0.272 & 0.637 & 0.341 \\
BERT-base & 0.272 & 0.693 & 0.408 \\
RRF (Initial + BERT-base)& \textbf{0.355}$^\blacktriangle$ & \bf 0.787$^\blacktriangle$ & \textbf{0.476}$^\blacktriangle$ \\ 
Oracle  & 0.754 & 0.956  & 0.926 \\ 
\hline
TREC-top-auto & 0.267 & 0.715 & 0.436 \\
TREC-top-manual & 0.405 & 0.879 & 0.589 \\ \hline 
\end{tabularx}
\vspace{-4mm}
\end{table}
In this experiment, we study the effect of QuReTeC when incorporated in the reranking step (Section~\ref{sec:reranking}).
We keep the initial ranker fixed for all QuReTeC models.
Table~\ref{tab:reranking} shows the results.
First, we see that BERT-base improves over the initial retrieval model that uses QuReTeC for query resolution on the top positions (second line).
Second, when we fuse the ranked listed retrieved by BERT-base and the ranked list retrieval by the initial retrieval ranker using RRF, we significantly outperform BERT-base on all metrics (third line).
This shows that the two rankers can be effectively combined with RRF, which is a very simple fusion method that only has one parameter which we do not tune.
We also see that our best model outperforms TREC-top-auto on all metrics.
Furthermore, by comparing RRF (line 3) to Oracle (line 4) we see that there is still plenty of room for improvement for reranking, which is a clear direction for future work.
This also shows that the TREC CAsT dataset is sufficiently challenging for comparing automatic systems.
Note that TREC-top-manual uses the gold standard query resolutions and is thereby not directly comparable with the rest of the methods. 
\subsection{Distant supervision for query resolution}
\label{sec:distant-supervision-query-resolution}
In this section we answer (RQ2): Can we use distant supervision to reduce the amount of human-curated query resolution data required to train QuReTeC?
\begin{figure}[t]
    \centering
  \centering
  \includegraphics[width=0.9\linewidth]{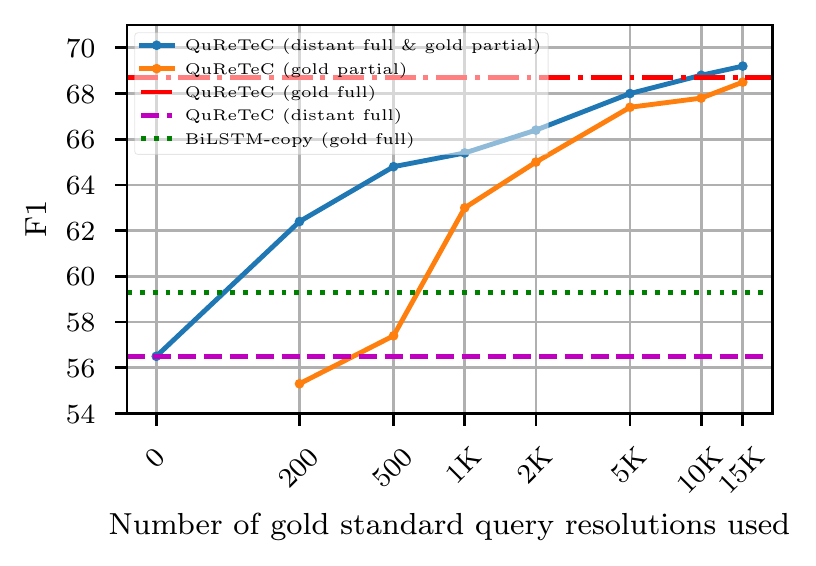}  
\caption{Query resolution performance (intrinsic) on the QuAC test set on different supervision settings. 
Gold refers to the QuAC train (gold) dataset and distant refers to the QuAC train (distant) dataset. 
Full refers to the whole and partial refers to a part of the corresponding dataset (gold or distant).  
The x-axis is plotted in log-scale.}
\label{fig:dist_sup_percentage_f1}

\end{figure}
Figure~\ref{fig:dist_sup_percentage_f1} shows the query resolution performance when training QuReTeC under different settings (see figure caption for a more detailed description).
For QuReTeC (distant full \& gold partial) we first pretrain QuReTeC on distant and then resume training with different fractions of gold. 
First, we see that QuReTeC performs competitively with BiLSTM-copy even when it does not use any gold resolutions (distant full).\footnote{Also, when trained with distant full, QuReTeC performs better than an artificial method that uses the label of the distant supervision signal as the prediction in terms of F1 (56.5 vs 41.6). This is in line with previous work that successfully uses noisy supervision signals for retrieval tasks~\cite{dehghani_neural_2017,voskarides-weakly-supervised-2018}.}
More importantly, when only trained on distant, QuReTeC performs remarkably well in the low data regime.
In fact, it outperforms BiLSTM-copy (trained on gold) even when using a surprisingly low number of gold standard query resolutions (200, which is $\sim$1\% of gold).
Last, we see that as we add more labelled data, the effect of distant supervision becomes smaller. This is expected and is also the case for the model trained on QuAC train (gold).\footnote{In fact (not shown in Figure~\ref{fig:dist_sup_percentage_f1}), performance stabilizes after 15K query resolutions ($\sim$75\% of gold full).}
In order to test whether our distant supervision method can be applied on different encoders, we performed an additional experiment where we replaced BERT with a simple BiLSTM as the encoder in QuReTeC.
Similarly to the previous experiment, we observed a substantial increase in F1 when retraining with 2K gold standard resolutions (+12 F1) over when only using gold resolutions.
In conclusion, our distant supervision method can be used to substantially decrease the amount of human-curated training data required to train QuReTeC.
This is especially important in low resource scenarios (e.g. new domains or languages), where large-scale human-curated training data might not be readily available.

\subsection{Analysis}
In this section we perform analysis on QuReTeC when trained with gold standard supervision.
\subsubsection{Query resolution performance per turn}
\begin{figure}[t]
\captionsetup[subfigure]{aboveskip=-1pt,belowskip=-1pt}
    \centering
    \begin{subfigure}{.5\linewidth}
  \centering
  \includegraphics[width=1.0\linewidth]{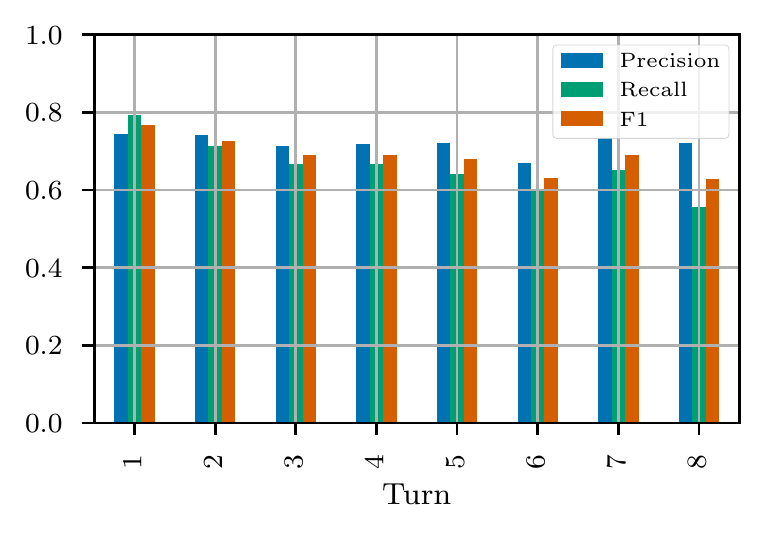}  
  \caption{QuAC \emph{test}}
  \label{fig:tc_per_turn_quac}
\end{subfigure}
~
\begin{subfigure}{.5\linewidth}
  \centering
  \includegraphics[width=1.0\linewidth]{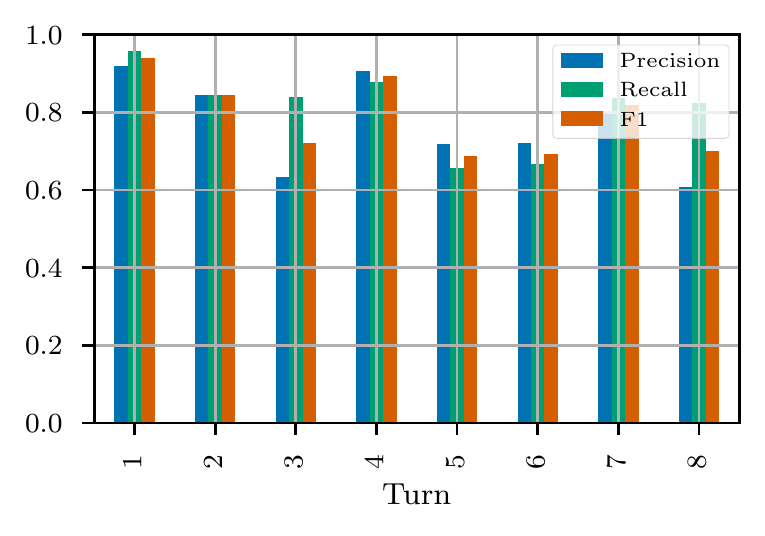}  
  \caption{CAsT \emph{test}}
  \label{fig:tc_per_turn_cast}
\end{subfigure}
\caption{Intrinsic query resolution evaluation (term classification performance) for QuReTeC, averaged per turn.}
\label{fig:tc_per_turn}
\end{figure}
Here we answer (RQ3) by analyzing the robustness of QuReTeC at later conversation turns.
We expect query resolution to become more challenging as the conversation history becomes larger (later in the conversation).
\textbf{Intrinsic} Figure~\ref{fig:tc_per_turn} shows the QuReTeC performance averaged per turn on the QuAC and CAsT datasets.
Even though performance decreases towards later turns as expected, we observe that it decreases very gradually, and thus we can conclude that QuReTeC is relatively robust across turns.

\textbf{Extrinsic  -- initial retrieval} 
\begin{figure}[t]
\captionsetup[subfigure]{aboveskip=-1pt,belowskip=-1pt}
    \centering
       \begin{subfigure}{.5\linewidth}
  \centering
  \includegraphics[width=\textwidth]{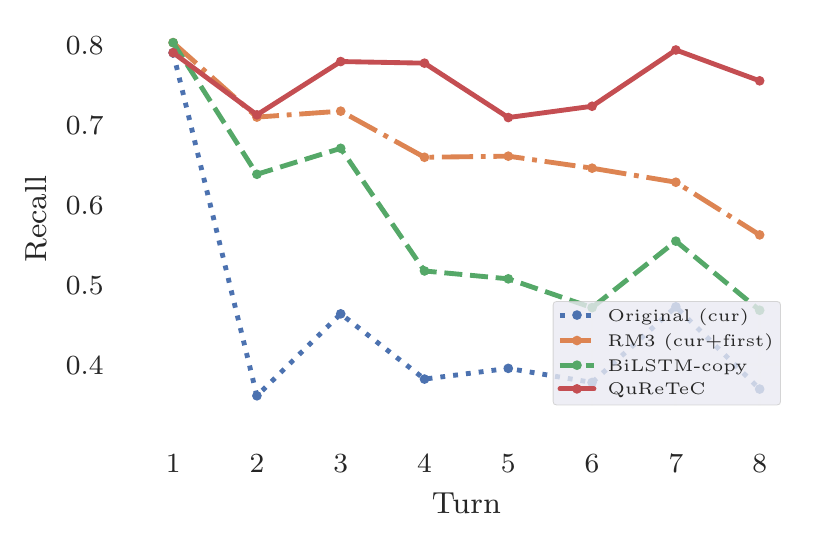}  
  \caption{Recall}
  \label{fig:initial_retrieval_per_turn_recall}
\end{subfigure}
~
    \begin{subfigure}{.5\linewidth}
  \centering
  \includegraphics[width=\textwidth]{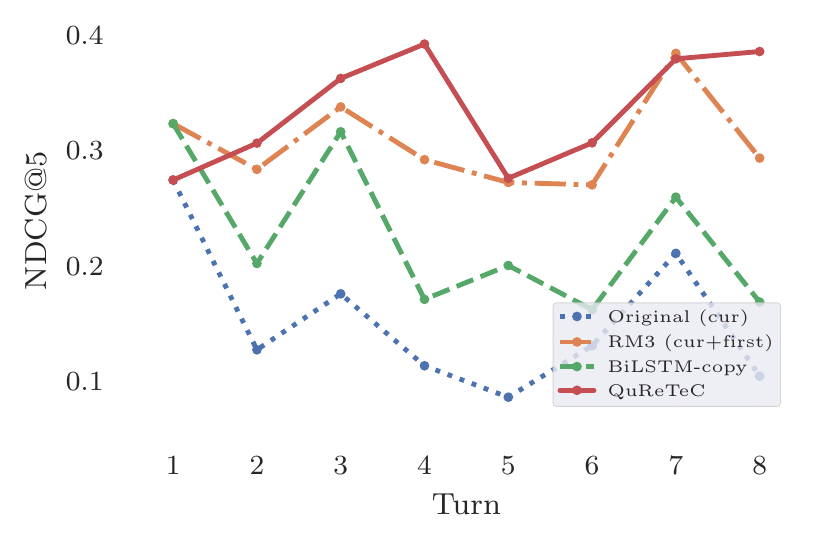}  
  \caption{NDCG@5}
  \label{fig:initial_retrieval_per_turn_recall_ndcg_5}
\end{subfigure}
\caption{Initial retrieval performance per turn for different query resolution methods CAsT \emph{test}
}
\label{fig:initial_retrieval_per_turn}
\end{figure}
Figure~\ref{fig:initial_retrieval_per_turn} shows the performance of different query resolution methods when incorporated in the initial retrieval step.
We observe that QuReTeC is robust to later turns in the conversation, whereas the performance of all the baseline models decreases faster (especially in terms of recall).
For reranking, we observe similar patterns as with initial retrieval; we do not include those results for brevity.

\subsubsection{Qualitative analysis}
Here we perform qualitative analysis by sampling specific instances from the data.
\begin{table}[t]
\centering
\small
\caption{Qualitative analysis for QuReTeC on query resolution (intrinsic). 
We denote true positive terms with underline and false negative terms in italics. The examples are sampled from the QuAC dev set.}
\label{tab:qualitative-intrinsic}
\begin{tabularx}{\linewidth}{p{0.97\linewidth}}
\hline
\textbf{Success case} -- no mistakes \\ 
\hline
Q1: What was \underline{Bipasha} \underline{Basu}'s debut?\\
A1: In 2001, Basu finally made her debut opposite Akshay Kumar in Vijay Galani 's \underline{Ajnabee}.\\
Q2: Did this help her become well known?\\
A2: It was a moderate box-office success and attracted unfavorable reviews from critics.\\
Q3 (current): Why did she receive unfavorable reviews? \\
\hline
\textbf{Failure case} -- misses two relevant terms: \emph{dehusking}, \emph{machine}\\
\hline
Q1: How old was \underline{Alexander} \underline{Graham} \underline{Bell} when he made his first invention?\\
A1: The age of 12.\\
Q2: What did he invent?\\
A2: Bell built a homemade device that combined rotating paddles with sets of nail brushes.\\
Q3: What was it for?\\
A3: A simple \emph{dehusking} \emph{machine}.\\
Q4 (current): By inventing this, what happened to allow him to continue inventing things? \\
\hline
\end{tabularx}
\end{table}
\textbf{Intrinsic}
Table~\ref{tab:qualitative-intrinsic} shows one success and one failure case for QuReTeC from the QuAC dev set.
In the success case (top) we observe that QuReTeC succeeds in resolving ``she'' $\rightarrow$ \{``Bipasha'', ``Basu''\} and  ``reviews'' $\rightarrow$ ``Anjabee''. Note that ``Anjabee'' is a movie in which Basu acted but is not mentioned explicitly in the current turn.
In the failure case (bottom) we observe that QuReTeC succeeds in resolving ``him'' $\rightarrow$ \{``Alexander'', ``Graham'' ``Bell''\} but misses the connection between ``this'' and ``dehusking machine''.
\begin{table}[t]
\centering
\small
\caption{Qualitative analysis for initial retrieval (extrinsic) when using QuReTeC or RM3 (cur+first) for query resolution. The example is sampled from the TREC CAsT dataset.}
\label{tab:qualitative-extrinsic}
\begin{tabularx}{\linewidth}{p{0.97\linewidth}}
\hline
Q1: What is a real-time database?\\
Q2: How does it differ from traditional ones?\\
Q3: What are the advantages of real-time processing?\\
Q4: What are examples of important ones?\\
Q5: What are important applications?\\
Q6: What are important cloud options?\\
Q7: Tell me about the Firebase DB? \\
Q8 (current): How is it used in mobile apps?\\
 \hline
\textbf{Predicted terms -- QuReTeC}: \{``database'', ``firebase'', ``db'' \} \\
\textbf{Top-ranked passage -- QuReTeC} \\
Firebase is a mobile and web application platform \ldots Firebase's initial product was a realtime database, \ldots Over time, it has expanded its product line to become a full suite for app development \ldots \\ \hline
\textbf{Predicted terms -- RM3 (cur+first)}: \{``real'', ``time'', ``database''\} \\ 
\textbf{Top-ranked passage -- RM3 (cur+first)} \\
There are two options in Jedox to access the central OLAP database and software functionality on mobile devices: 
Users can access reports through the touch-optimized Jedox Web Server \ldots
 on their smart phones and tablets.\\
\hline
\end{tabularx}
\vspace{-4mm}

\end{table}
\textbf{Extrinsic -- initial retrieval}
Table~\ref{tab:qualitative-extrinsic} shows an example from the CAsT test set where QuReTeC succeeds and RM3 (cur+first), the best performing baseline for initial retrieval, fails.  
First, note that a topic change happens at Q7 (the topic changes from general real-time databases to Firebase DB). 
We observe that QuReTeC predicts the correct terms, and a relevant passage is retrieved at the top position.
In contrast, RM3 (cur+first) fails to detect this topic change and therefore an irrelevant passage is retrieved at the top position that is about real-time databases on mobile apps but not about Firebase DB.
%

%
\section{Conclusion}

In this paper, we studied the task of query resolution for conversational search.
We proposed to model query resolution as a binary term classification task: whether to add terms from the conversation history to the current turn query.
We proposed QuReTeC, a neural query resolution model based on bidirectional transformers.
We proposed a distant supervision method to gather training data for QuReTeC.
We found that QuReTeC significantly outperforms multiple baselines of different nature and is robust across conversation turns.
Also, we found that our distant supervision method can substantially reduce the required amount of gold standard query resolutions required for training QuReTeC, using only query-passage relevance labels. 
This result is especially important in low resource scenarios, where gold standard query resolutions might not be readily available.
As for future work, we aim to develop specialized rankers for both the initial retrieval and the reranking steps that incorporate QuReTeC in a more sophisticated way.
Also, we want to study how to effectively combine QuReTeC with text generation query resolution methods as well as pseudo-relevance feedback methods.
Finally, we aim to explore weak supervision signals for training QuReTeC~\cite{dehghani_neural_2017}.

\begin{acks}
We thank the anonymous reviewers for their feedback.
This research was partially supported by the Netherlands Organisation for Scientific Research (NWO) under project nr CI-14-25,
the NWO Innovational Research Incentives Scheme Vidi (016.Vidi.189.039),
the NWO Smart Culture - Big Data / Digital Humanities (314-99-301),
the H2020-EU.3.4. - SOCIETAL CHALLENGES - Smart, Green And Integrated Transport (814961),
the Google Faculty Research Awards program,
and the Innovation Center for AI (ICAI).
All content represents the opinion of the authors, which is not necessarily shared or endorsed by their respective employers and/or sponsors.
\end{acks}

\section*{Code and data}
To facilitate reproducibility, we share the resources used in this paper at \url{https://github.com/nickvosk/sigir2020-query-resolution}.

\bibliographystyle{ACM-Reference-Format.bst}
\bibliography{bibliography}

\end{document}